\documentstyle[epsf]{mn}
\newcommand{\etal}{{et al}\/.}
\def\dgr{\degr}
 
\begin{document}
\title[Dynamics of 3C\,449]{Dynamics of the radio galaxy 3C\,449}
\author[Hardcastle, Worrall and Birkinshaw]{M.J.\ Hardcastle$^1$,
D.M.\ Worrall$^{1,2}$ and M.\ Birkinshaw$^{1,2}$\\
$^1$Department of Physics, University of Bristol, Tyndall Avenue,
Bristol BS8 1TL\\
$^2$Harvard-Smithsonian Center for Astrophysics, 60 Garden Street,
Cambridge, MA 02138, U.S.A.}
\maketitle
\begin{abstract}
We present {\it ROSAT} PSPC observations of the twin-jet radio galaxy
3C\,449. The soft X-ray emission from this object is dominated by an
extended halo with a scale comparable to that of the radio source. The
asymmetry of the X-ray emission is reflected in that of the radio
lobes, providing evidence that the behaviour of the jets is strongly
influenced by the external medium. A region of reduced X-ray surface
brightness coincident with the southern radio lobe of 3C\,449 suggests
that the radio source has displaced thermal plasma from the
X-ray-emitting halo. However, the minimum pressure in the radio lobe
is considerably lower than our estimates of the pressure in the
external medium. We discuss the implications for the dynamics of the
radio source.
\end{abstract}

\section{Introduction}

3C\,449 (B2 2229+39) is a well-known FRI radio galaxy (Fanaroff \&
Riley 1974) at a redshift of 0.0171. Its symmetrical inner jets have
been well studied in the radio (Perley, Willis \& Scott 1979; Cornwell
\& Perley 1984). On larger scales, the southern jet flares into a
lobe, while the northern jet continues to be collimated until it fades
into the noise (Birkinshaw, Laing \& Peacock 1981; J\"agers 1987;
Andernach \etal\ 1992; Leahy, Bridle \& Strom 1997). 3C\,449's host
galaxy, UGC 12064, has a close bright companion (e.g.\ Balcells
\etal\ 1992) and models have been proposed in which the orbital motion
of the host is responsible for the symmetrical oscillations in the
jets (e.g.\ Lupton \& Gott 1982; Hardee, Cooper \& Clarke 1994).

UGC 12064 appears in the optical to lie in a poor cluster (Zw
2231.2+3732), and the extended optical halo surrounding it and its
companion allows it to be classed as a cD galaxy (Wyndham 1966). Miley
\etal\ (1983) used the {\it Einstein} IPC to show that it is a weak,
extended X-ray source, and the rotation measure gradients in the jet
are thought to be produced by the hot gas responsible for the X-ray
emission (Cornwell \& Perley 1984). In this paper we present new,
sensitive observations of 3C\,449 with the {\it ROSAT} PSPC, which
allow us to map in detail the hot gas halo surrounding the source.

Throughout we use a cosmology in which $H_0 = 50$ km s$^{-1}$
Mpc$^{-1}$, $q_0 = 0$. At the distance of 3C\,449, 1 arcsec
corresponds to 0.485 kpc.

\section{Observations and data analysis}
\label{obs}

3C\,449 was observed for 10 ks with the ROSAT PSPC between 1993 Jan 04
and 1993 Jan 10 as part of a programme to study the X-ray emission
from a sample of radio galaxies drawn from the B2 radio survey as
likely candidates for the parent population of BL Lac objects (Worrall
\& Birkinshaw 1994). We analysed the data with the Post-Reduction
Off-line Software (PROS). After filtering to remove time intervals
with abnormally high particle background, we were left with 9.15 ks of
good data.

An extended source was detected with $1840 \pm 85$ net counts in a
10-arcmin-radius circle about the pointing centre, corresponding to a
count rate of $0.20 \pm 0.01$ s$^{-1}$ over the PSPC energy range
(0.1--2.5 keV). The background count rate was estimated from an
annulus between 10.8 and 16.7 arcmin from the pointing centre; source
and background regions were corrected for the effects of vignetting
using the 1-keV parametrisation of the vignetting function. A number
of point X-ray sources, and one extended source, believed to be
background sources, were excluded from the source and background
regions. The closest of these, 8.6 arcmin W of the source, is
coincident with a foreground star; there is no bright optical object
coincident with the others, though a source 15 arcmin to the SE is
coincident with a weak radio source.

Fig.\ \ref{xrayc} shows the extended X-ray structure of the
source. Extended object and X-ray
background analysis software due to S.\ Snowden (Snowden 1997) was
used to model and subtract non-cosmic background components and
correct for exposure and vignetting. The resulting 0.4-2 keV image was
then adaptively smoothed as described elsewhere (Worrall, Birkinshaw \&
Cameron 1995; Worrall \etal\ in prep.). The central peak in
X-ray brightness corresponds to the position of the radio core of
3C\,449 to within 9 arcsec, consistent within the absolute position
errors of {\it ROSAT}. The central regions around the peak are clumpy
and elongated in position angle $\sim 45\dgr$ on scales of up to 10
arcmin (300 kpc). The reason for this elongation is not clear; there
are no bright associated galaxies coincident with it on sky-survey
plates. The X-ray emission on larger scales is asymmetrical, being
significantly more extended to the south and west than it is to the
north and east.

Spectral fits to the data were carried out in the energy range
0.2--1.9 keV (SASS channels 6--29 inclusive), for consistency with
earlier work and because this is the energy band over which the PSPC
point-response function can be modelled accurately. A thermal model is
a better fit to the data than a power law, as expected from the
extended nature of the X-ray emission. The best-fitting single thermal
model at the redshift of the host galaxy has abundances 100 per cent
solar, $kT = 1.21_{-0.06}^{+0.14}$ keV, galactic $N_H = 5^{+6}_{-3}
\times 10^{20}$ cm$^{-2}$, with errors being $1\sigma$ for two
interesting parameters. This value of $N_H$ is consistent with the
value interpolated from the results of Stark \etal\ (1992), $1.1
\times 10^{21}$ cm$^{-2}$, at the $1\sigma$ level; fixing galactic
$N_H$ at this value and allowing intrinsic $N_H$ to vary gives a
best-fit model with 50 per cent solar abundances and $kT =
1.14^{+0.09}_{-0.06}$ keV, with negligible intrinsic absorption. More
complicated models, such as a two-component thermal model or a thermal
model plus power law, do not improve the fit to the data; there is
also no significant difference in the temperatures fitted to the inner
and outer regions of the source consistent with little variation of
the gas temperature across the source. The count rate and spectrum
imply a 0.2--1.9 keV luminosity of $4.2 \times 10^{35}$ W; the source
is thus comparable to the brightest sources studied by Worrall \&
Birkinshaw (1994).

Although the source is not azimuthally symmetrical, radial profile
fitting allows its size to be roughly characterised. We therefore
extracted a background-subtracted radial profile as described by
Worrall \& Birkinshaw (1994), and determined the best-fitting model
consisting of a $\beta$-model (Sarazin 1986) and an unresolved
component. The best-fitting models have low $\beta$; with $\beta =
0.35$ the best-fit model had a core radius of 35 arcsec, with a
central gas density $4.6 \times 10^3$ m$^{-3}$. The radial profile and
best-fit model are plotted in Fig.\ \ref{radial}. The
unresolved component contained $\sim 2$ per cent of the counts; any
nuclear component is therefore weak in comparison to the extended
emission, as suggested by the spectral-model fits. The particular
value of $\beta$ chosen does not strongly affect the results discussed
in the remainder of the paper.  Models with $\beta
\leq 2/3$ formally require a cutoff radius to prevent the gas mass
becoming infinite, but, as noted by Birkinshaw \& Worrall (1993), the
effect on the model emission measure distribution is small over the
region in which the X-ray-emitting gas is detected.

\section{Radio-X-ray comparison}

Fig.\ \ref{overlay} shows the 608-MHz WSRT map of Leahy \etal\ (1997)
overlaid on a grey-scale representation of the image of Fig.\
\ref{xrayc}. The southern lobe appears to be embedded in X-ray
emitting material, while the northern plume is clear of it after the
bend at 5 arcmin (150 kpc) from the core. This may be interpreted as
evidence for interaction between the outflow from the radio source and
the ambient medium; the differences between the southern lobe and
northern plume are then attributed to the different external
environments, with the higher inferred densities to the south being
sufficient to confine the outflow and form a lobe.

Further evidence suggesting that the radio emission is interacting
with the X-ray-emitting medium (rather than just being projected on to
it) comes from an apparent local surface-brightness minimum in the
X-ray emission coincident with the southern lobe; this suggests that
the pressure from the relativistic particles in the lobe has displaced
the thermal plasma, which now seems to be forming a rim around it. We
emphasise, however, that the individual features forming the southern
edge of the rim are detected, on maps smoothed with a Gaussian with
$\sigma=1$ arcmin, at the 3--$5\sigma$ level only; further
observations are needed to confirm the nature of these features. (We
discuss the significance of features in smoothed X-ray images in
Appendix A.) Fig.\ \ref{slice} shows the edge-brightened nature of
this rim; the background-corrected surface brightness drops by more
than a factor of 2 between edge and minimum, although the cut of Fig.\
\ref{slice} passes through an off-centre blob of X-ray emission. When
mean count rates are measured from the vignetting-corrected 0.1-2.5
keV data in the deficit region (defined using the radio contours as a
guide) and the `rim' (defined as a region extending 2-3 arcmin out
from the deficit region, but excluding the bright central regions of
the source), the mean count rate (source and background) in the rim
[$(4.7 \pm 0.1) \times 10^{-7}$ counts arcsec$^{-2}$ s$^{-1}$] exceeds
that in the deficit region [$(3.8 \pm 0.2) \times 10^{-7}$ counts
arcsec$^{-2}$ s$^{-1}$] at the $3\sigma$ level, so we regard the
deficit of X-ray emission as significant. Similar X-ray deficits
associated with radio lobes have been seen in the cluster-centre
galaxy NGC 1275 by B\"ohringer \etal\ (1993) and in the powerful FRII
radio galaxy Cygnus A (Carilli, Perley \& Harris 1994). We cannot rule
out the possibility that the radio lobe coincides with the X-ray
minimum purely by chance, but the coincidence in size between the
radio lobe and the spacing of the `rim' features is certainly
suggestive.

The inner and outer regions of 3C\,449 have different rotation
measures (RM); Cornwell \& Perley (1984) measure -210 rad m$^{-2}$
near the source centre while Andernach \etal\ (1992) find RM $\sim
-170$ rad m$^{-2}$ in the outer lobes. The bulk of the RM is probably
contributed by our Galaxy; nearby sources have average RM $\sim -210$
rad m$^{-2}$, though the situation is complicated by a foreground HII
region which crosses 3C\,449 (Andernach \etal ). If the difference
between the central and outer regions of 3C\,449 is due to the cluster
gas, then it can be produced with uniform magnetic fields in the
X-ray emitting gas of roughly 10 pT.

\section{Discussion: dynamics of the radio source}

If the radio-emitting material has displaced thermal plasma then we
expect pressures to be comparable in the southern lobe and surrounding
gas. Figure \ref{pressure} shows the pressure predicted by the
best-fitting $\beta$-model (with $\beta=0.35$, core radius 35 arcsec
and $kT = 1.14$ keV) as a function of distance from the source centre,
together with minimum pressures at various points in the radio
source. It can be seen that if the radio source is in the plane of the
sky the minimum pressures are lower than the inferred hot gas pressure
at all points in the source, with the discrepancy being greatest (a
factor of 20) at the far ends of the object. Similar results are
obtained by using other models for the ambient gas: for example, if
the southern region of the X-ray source is modelled as a uniform
spherical shell with inner radius 4.5 arcmin and outer radius 6.2
arcmin (Fig.\ \ref{slice}), neglecting the off-centre blob of emission
which may be a foreground or background feature, the thermal pressure
is $1.7 \times 10^{-13}$ Pa, slightly higher than the value predicted
by the $\beta$-model.

Underpressuring of the radio lobes with respect to the
X-ray-emitting plasma has been observed in other low-power sources
(Morganti \etal\ 1988; Killeen, Bicknell \& Ekers 1988; Feretti,
Perola \& Fanti 1992; B\"ohringer \etal\ 1993; Worrall \etal\
1995). We can attempt to resolve the discrepancy by considering
projection effects on the radio source, by considering dynamical
pressure balance, by allowing relativistic protons to make a
significant contribution to the energy density, by considering
situations where the energy densities are far from equipartition, or
by allowing internal thermal material. These possibilities will be
discussed in turn.

The assumption that the source is in the plane of the sky may be
incorrect. If the source makes an angle to the line of sight $\theta$
both the measured distance from the centre and the length used in
minimum-energy calculation are underestimated by a factor $1/\sin
\theta$; the minimum pressure in a given component drops as a result
by a factor of roughly $(\sin \theta)^{4/7}$. Because of this, small
angles to the line of sight ($\la 8\dgr$) are necessary to achieve
pressure balance even in the inner regions of the source where the
pressures are most nearly matched. This is implausible both because of
the observed symmetry of the jets (the degree of symmetry seen in
high-resolution VLA radio maps would require velocities in the jets at
5 kpc from the core of $\la 0.07c$, compared with the estimated
velocities of $\ga 0.4c$ at that distance from the core in similar
objects: Hardcastle \etal\ 1996, 1997; Laing 1996) and because of the
implausibly large true linear sizes ($\ga 5$ Mpc) they would imply for
the radio source. This is unlikely to provide a solution to the
problem, therefore.  If we use the velocities inferred from
observations of other sources, the angle to the line of sight $\theta$
is predicted from the jet sidedness to be $\sim 80\dgr$.

Pressure balance could in principle be dynamical if the lobe were
expanding; if we postulate a spherically symmetric wind, driven by the
jet, impinging on the external medium, then the problem is similar to
that of the hot spots of classical double radio sources, and the
non-relativistic equation of pressure balance in the frame of the edge
of the lobe is
\[
P_{\rm int} + \rho_{\rm int}(v_{\rm wind}-v_{\rm exp})^2 = P_{\rm ext}
+ \rho_{\rm ext}v_{\rm exp}^2
\]
where $v_{\rm exp}$ is the outward velocity of the lobe's edge. This
can be rewritten in terms of the internal sound speed $a_{\rm int}$
and Mach number $\cal M$ of the wind:
\[
{\cal M}^2a_{\rm int}^2 = \left ({{P_{\rm ext}}\over{P_{\rm
int}}}-1\right ){{a_{\rm int}^2}\over{\Gamma}} + 2{\cal M}a_{\rm
int}v_{\rm exp} + \left({{\rho_{\rm ext}}\over{\rho_{\rm
int}}}-1\right)v_{\rm exp}^2 \]
(where $\Gamma$ is the internal adiabatic index) from which it can be
seen that no subsonic solution exists if $\rho_{\rm ext} > \rho_{\rm
int}$ and $P_{\rm ext} \gg P_{\rm int}$. For a purely relativistic
internal plasma $a_{\rm int} \sim c/{\sqrt 3}$, so a relativistic
wind would be required for pressure balance, and we would expect to
see effects of beaming on the surface brightness of the lobe. If cold
denser material were mixed into the plasma (e.g.\ entrained material,
discussed below) then $a_{\rm int}$ would drop and $P_{\rm int}$ and
$\rho_{\rm int}$ would rise, making lower wind speeds possible. The
energetics of this process, though demanding, are not completely
inconsistent with the power that could be supplied by the jet. 
It is harder to apply this explanation to the underpressuring in the
inner regions of the source.

The minimum pressure of the radio plasma is a weak function of the
ratio $\kappa$ between the number of radiating and non-radiating
relativistic particles in the plasma, going roughly as $(1 +
\kappa)^{4/7}$. A large contribution from relativistic protons
($\kappa \sim 200$) is necessary before the internal and external
pressures in the outer regions of the source become similar. A plasma
filling factor $\phi$ of less than $1/200$ has the same
effect. Neither of these situations can be ruled out. Filamentary
structures seen in radio maps of the best-studied objects at high
resolution may indicate filling factors of this order when projection
effects are taken into account.  In the inner jets, where the pressure
discrepancy is less, a more moderate contribution from protons
($\kappa \sim 20$) or a larger filling factor ($\phi \sim 1/20$) is
sufficient to achieve pressure balance.

The electron and magnetic field energy densities may be far from
equipartition. The equipartition $B$-field in the southern lobe,
treated as a uniform sphere with $\kappa = 0$, is approximately 0.14
nT; to achieve the required pressures, the $B$-field must be either
0.9 nT (a factor of 6 above equipartition) or 30 pT (a factor of 5
below equipartition). In the latter case, the flux in X-rays from
inverse-Compton scattering of the cosmic background radiation
(calculated using the code of Hardcastle, Birkinshaw \& Worrall 1998)
would be significant and detectable with our observations, so this
possibility can be neglected and only a high $B$-field is consistent
with observation. However, Feigelson \etal\ (1995) report on X-ray
observations of Fornax A which are consistent with a lobe magnetic
field slightly {\it weaker} than equipartition (with $\kappa = 0$,
$\phi=1$) in that object. Again, in the inner jets, the problem is
less severe; the $B$-field must be about a factor 4 above
equipartition, or a factor 3 below it.

The jets will entrain material as they pass through the galaxy (e.g.\
Bicknell 1994 and references therein), and the entrained gas will
provide a contribution to the internal pressure. The fact that the
discrepancy between internal and external pressures increases with
distance from the galaxy is consistent with this model, since we would
expect the amount of entrained material to increase in this way. But
the fact that there is a deficit in X-ray emission in the S lobe
suggests that there is little hot ($kT \sim 1$ keV) gas in this
region, and so that there is little contribution to the internal
pressure from hot gas. Cold gas would not be observed in X-rays but
would contribute less to $P_{\rm int}$. The observed depolarization
between $\lambda=49$ and 21 cm, (DP$\sim 0.75$; J\"agers 1987) in the
centre of the lobe suggests that the internal thermal electron density
$n_e \la 60$ m$^{-3}$, using the simple models of Cioffi \& Jones
(1980), which is less than the inferred external particle density at
this radius ($n \sim 300$ m$^{-3}$). Depolarization, at least in the
inner regions of the source, can be attributed entirely to rotation
measure gradients across it (Cornwell \& Perley 1984). However, field
reversals can hide large quantities of thermal material (Laing 1984),
so it is possible that thermal material at temperatures significantly
different from 1 keV contributes to the internal pressure; the fact
that the 608-MHz polarization is much higher at the edges of the lobe
than at the centre is certainly suggestive of internal depolarization.

If the radio source is genuinely underpressured with respect to the
hot gas, the association between the radio lobe and the X-ray void
must be considered coincidental, in the sense that the lobe has simply
filled a pre-existing region of low gas density. This is not
completely out of the question, since the collapse time of such a
region is comparable to $D/c_{\rm X}$, where $D$ is the size of the
region and $c_{\rm X} (\sim 400\ {\rm km\ s}^{-1})$ is the sound speed
in the X-ray gas; this time is a few $\times 10^8$ years, comparable
to the lifetime of the radio source. The hypothesis is consistent with
the generally unrelaxed appearance of the X-ray emission. The problem
again is to account for the underpressuring in the inner regions, where
the collapse times are much shorter.

\section{Conclusion}

We have presented evidence that the large-scale structure of 3C\,449 is
being determined by the distribution of the hot intergalactic
plasma.

The radio lobes are significantly underpressured if the radio-emitting
plasma is electron/positron with filling factor unity and the magnetic
field and particles are in energy equipartition. It is then hard to
understand how the southern lobe can be associated with a deficit of
X-ray emission. This problem, which
seems to be common in low-power radio sources, may be resolved in a
number of ways; the most plausible seem to be a dominant contribution
to the particle energy density from relativistic or thermal (entrained)
protons or a plasma filling factor much less than unity.

\section*{ACKNOWLEDGEMENTS}

Support from PPARC grant GR/K98582 and NASA grant NAG 5-1882 is
gratefully acknowledged. We are grateful to Larry Rudnick for
providing us with his 5-GHz VLA map of 3C\,449 and to the referee, Dr
Leahy, for drawing the polarization in the maps of J\"agers (1987) to
our attention, for discussion of the statistics in Appendix A, and for
a number of other helpful comments.

\appendix
\section{Uncertainties in smoothed X-ray images}

It is often necessary to convolve X-ray images with one or more smoothing
Gaussians in order to analyse extended structure. The disadvantage of
this procedure is that it renders a calculation of the significance of
detected structure or of the uncertainties on measured quantities more
difficult. Here we outline a means of overcoming the problem.

When a two-dimensional array of Poisson noise, with mean $\mu$ counts
per pixel, is convolved with a Gaussian ($f(r) \propto
\exp(r^2/2\sigma^2)$) it is tempting, but incorrect, to treat the
Gaussian as a single bin of size $2\pi \sigma^2$ and apply Poisson
statistics (with a mean of $2\pi \sigma^2 \mu$) to calculate the
significance of the results. This approach badly overestimates the
spread of the cumulative probability distribution (CPD), and so
overestimates the errors, because it treats the value of
each pixel in the convolved array as the uniformly weighted sum of a
small, finite number of single pixels containing Poisson-distributed
values, rather than as the Gaussian-weighted sum of a large (formally
infinite) number of such pixels.

A more rigorous treatment produces a better answer. Let the number of
counts in a pixel in a smoothed image, $c_j$, be the weighted sum of
the contributions from the pixels in the original image, $c_i$; $c_j =
\sum w_i c_i$, where $w_i$ are the weights. If the expected number of
counts per pixel in the unsmoothed image was $\mu$, then $\langle c_j
\rangle = \mu \sum w_i$; clearly we require $\sum w_i = 1$. The
variance of $c_j$, is given by ${\rm Var}(c_j) = \sum w_i^2 {\rm
Var}(c_i)$; for Poisson-distributed $c_i$, ${\rm Var}(c_j) = \mu \sum
w_i^2$. If the smoothing function is a 2-D Gaussian adequately sampled
by the pixels, then we can write $w_i = C \exp(-r/2\sigma^2)$,
where the condition $\int w_i 2\pi r {\rm d}r = 1$ requires that $C =
1/2\pi \sigma^2$; then
\[
{\rm Var}(c_j) = \mu C^2 \int_0^\infty \exp(-r/\sigma^2) 2\pi r {\rm
d}r = {\mu\over{4\pi\sigma^2}} \] so that the standard deviation of
the distribution is $\sqrt{\mu/4\pi\sigma^2}$. This result differs by
a factor of only $\sqrt{2}$ from that derived from the naive approach
discussed above. In the limit that $\sigma^2\mu \gg 1$, this
calculation therefore allows an accurate calculation of the significance of a
detection; the Central Limit Theorem guarantees that the cumulative
probability distribution (CPD) approaches normality, so that the
statistics can be treated as Gaussian, and the large $\sigma$ means
that the assumption of a well-sampled convolving function is
valid. However, when $\sigma^2\mu \la 1$, these assumptions are not
applicable and application of the result systematically
underestimates the true errors.

Because of this problem, the most general approach, and the one we
have adopted in this paper, is to use numerical (Monte Carlo) methods
to determine the correct CPD. The simplest numerical route to the CPD
involves generating a large number of arrays of pixels containing
Poisson noise with the correct mean number of counts per pixel $\mu$,
convolving with the Gaussian of interest, and generating an
approximate CPD from the results. This can then be used to find the
number of counts per convolved pixel corresponding to the 1, 2 and
3-$\sigma$ probability levels of a normal distribution. Results from
this procedure are plotted in Fig.\ \ref{gauss}. As expected, the CPD
converges on a normal distribution as the Gaussian size $\sigma$ tends
to infinity and/or for large $\mu$; consequently, the standard
deviation of the smoothed dataset is a good estimator of significance
level.  However, in the more usual case in X-ray astronomy where the
Gaussian has a $\sigma$ of a few pixels and $\mu \ll 1$ the noise is
far from normally distributed, and the standard deviation of the
convolved array is {\it not} a good estimator of the 68 per cent
confidence (`$1\sigma$') level; Monte Carlo methods then seem to be
required. The approach can easily be extended to calculating the CPD
of a binned array of smoothed Poisson noise, and automatically takes
account of the fact that nearby pixels are not independent after
smoothing. This method was used to find the error bars in Fig.\
\ref{slice}.

\bsp
\clearpage
\renewcommand{\thefigure}{\arabic{figure}}
\begin{figure*}
\begin{center}
\leavevmode
\vbox{\epsfysize 12cm\epsfbox{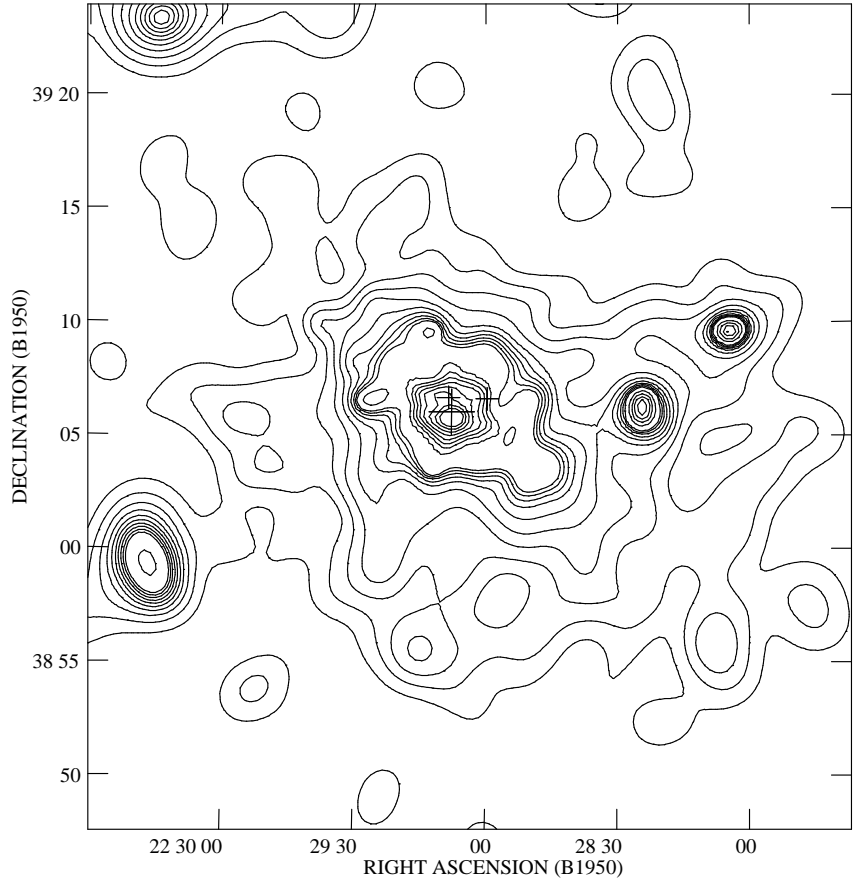}}
\caption{Background-subtracted X-ray contours of the extended emission around
3C\,449 in the PSPC energy band 0.4-2 keV (SASS channels 13 to 30
inclusive), adaptively smoothed as described in the text. Contour
levels (above the background) are $0.15 \times 10^{-3} \times (1, 2,
3, \dots, 10, 15, 20,\dots, 50)$ counts arcmin$^2$ s$^{-1}$. (The
corresponding negative contours enclose no data and so do not appear.)
The cosmic-background contribution was estimated to be $0.27 \times
10^{-3}$ counts arcmin$^2$ s$^{-1}$. The large cross marks the optical
position of 3C\,449 and the smaller crosses those of two nearby
galaxies.}
\label{xrayc}
\end{center}
\end{figure*}

\begin{figure*}
\begin{center}
\leavevmode
\vbox{\epsfysize 12cm\epsfbox{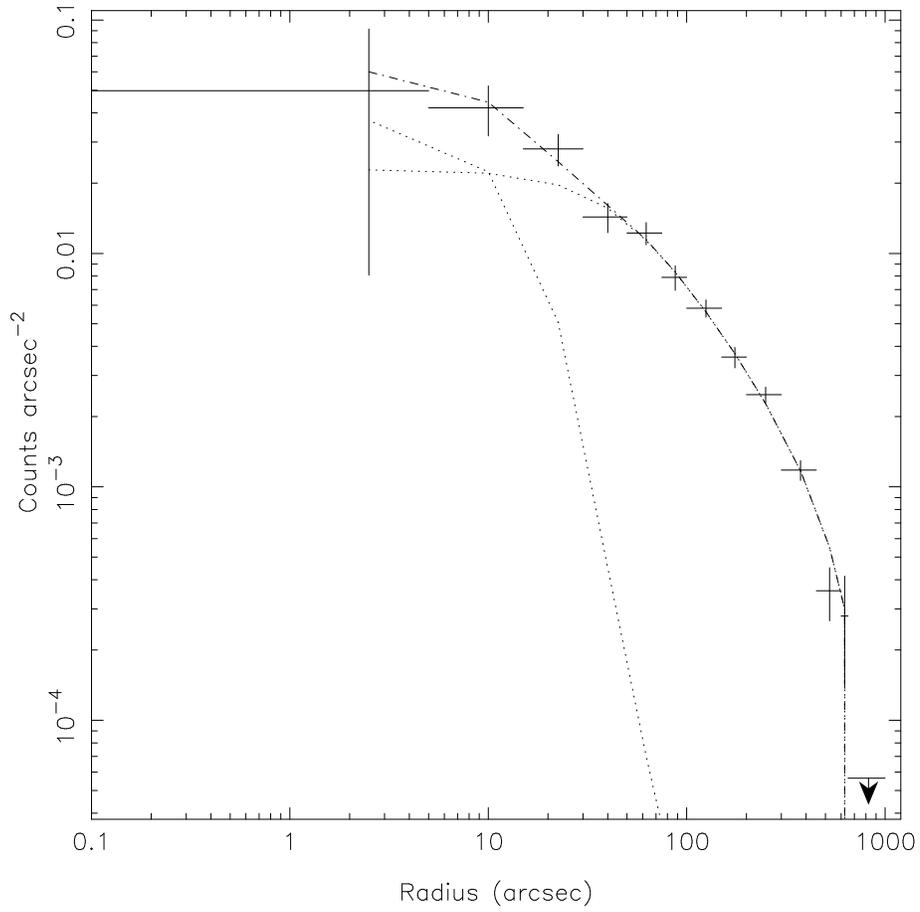}}
\caption{Background-subtracted radial profile of the X-ray emission
around 3C449 between 0.2 and 1.9 keV. The dotted lines show the two
components of the best-fit model and the dot-dashed line shows the
total model for comparison with the data. $\chi^2 = 7.7$ with 11
degrees of freedom.}
\label{radial}
\end{center}
\end{figure*}

\begin{figure*}
\begin{center}
\leavevmode
\vbox{\epsfysize 12cm\epsfbox{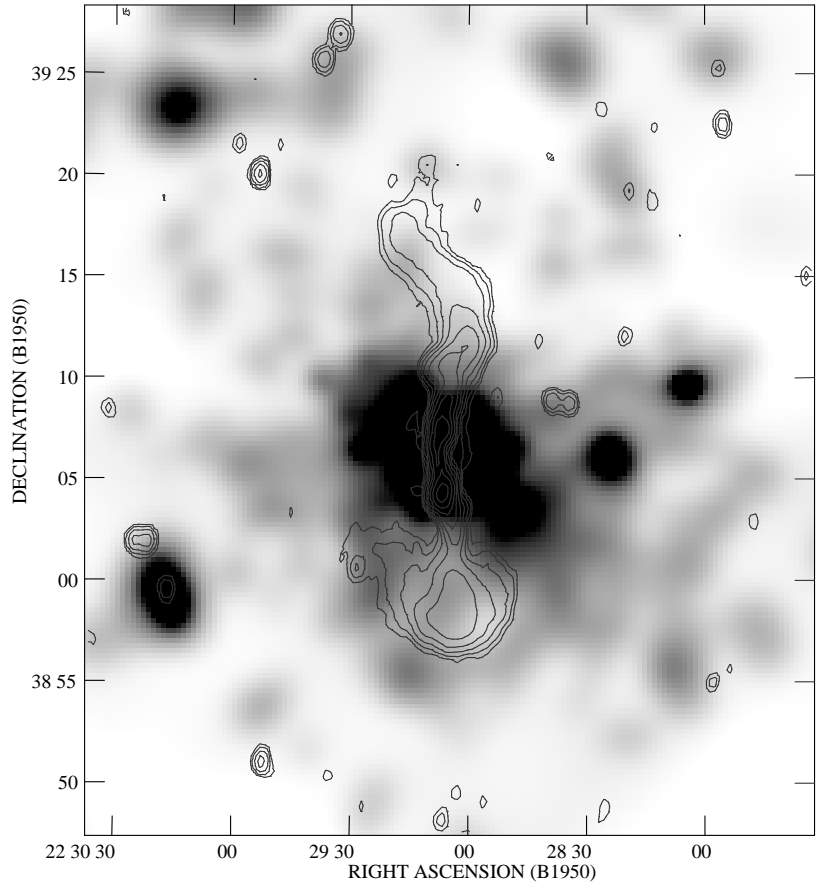}}
\caption{Radio contours at 608 MHz of 3C\,449 (at $2 \times (1, 2,
4\dots)$ mJy beam$^{-1}$; beam size $48 \times 30$ arcsec) superposed
on a greyscale representation of the image of Fig.\ \ref{xrayc}. Black
is at $1.2 \times 10^{-3}$ counts arcmin$^{-2}$ s$^{-1}$.}
\label{overlay}
\end{center}
\end{figure*}
\clearpage
\begin{figure}
\epsfxsize 8cm
\epsfbox{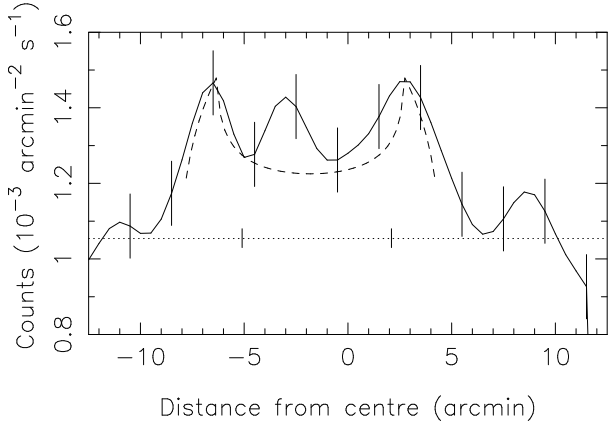}
\caption{East-west slice through the X-ray emission from 3C\,449 at
the widest point of the southern `ring' of X-ray emission, centred on
RA 22 28 54.4 DEC +38 58 54. West is to the left. 0.1-2.5 keV X-ray
data (all channels) binned into 15-arcsec pixels and smoothed with a
Gaussian with $\sigma = 1$ arcmin; each data point represents the
average over a bin with width 0.5 arcmin (measured east-west) and
length 2.5 arcmin (measured north-south). Superposed is the profile
expected from a spherical shell with inner radius 4.5 arcmin and outer
radius 6.2 arcmin, arbitrarily normalised. The dotted line shows the
background level measured from the background annulus described in
section \ref{obs}; bars crossing it mark the approximate size of the
southern radio lobe. Error bars, spaced at one per 4 data points, are
$\pm 1\sigma$ (68 per cent confidence level) for the given background
noise level, convolving Gaussian and binning scheme, derived from
simulation (see Appendix A).  The off-centre excess in the data is
interpreted in the text as foreground or background emission.}
\label{slice}
\end{figure}

\begin{figure}
\epsfxsize 8cm
\epsfbox{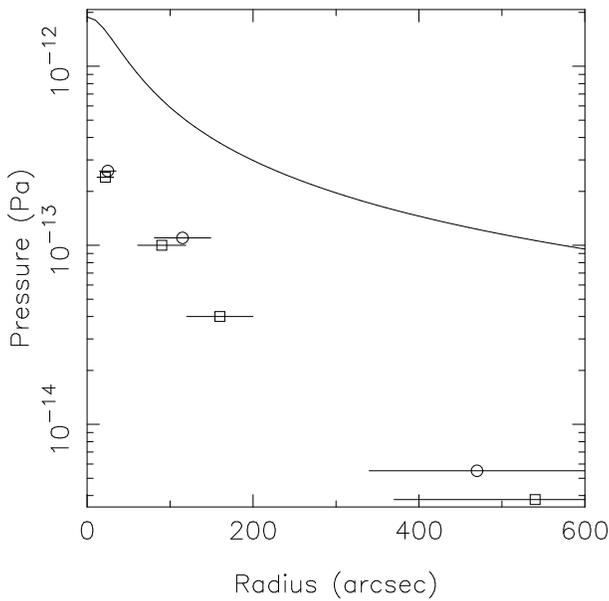}
\caption{Pressure from the hot gas as a function of distance from the
centre, calculated from the best-fit $\beta$-model with core
radius 35 arcsec and $\beta=0.35$. Superposed are measurements of
minimum pressure of various regions in the radio source. Squares
denote measurements made north of the core, circles those made to the south.}
\label{pressure}
\end{figure}

\appendix
\renewcommand{\thefigure}{A\arabic{figure}}
\begin{figure*}
\epsfxsize \linewidth
\epsfbox{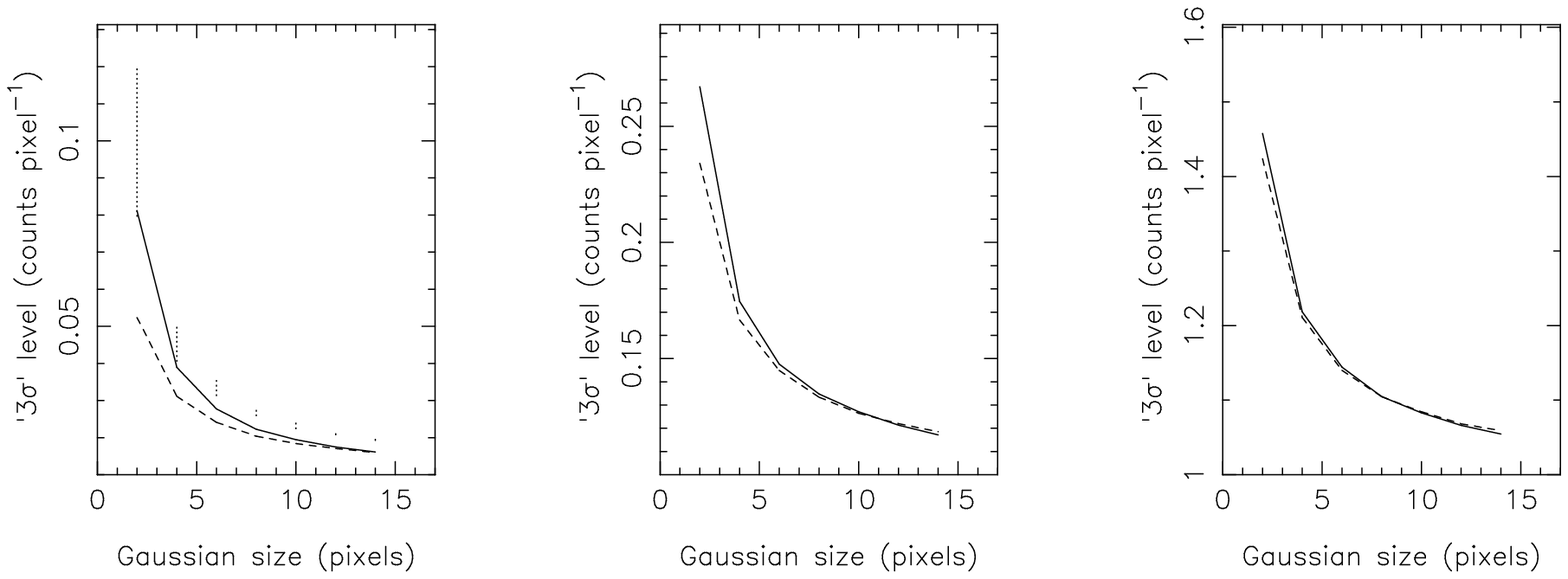}
\caption{Effects of different convolving Gaussians on the `$3\sigma$'
level of an array of initially Poisson-distributed noise. The level of
a contour that would exclude all but 0.135 per cent of the background
noise, in counts pixel$^{-1}$, is plotted (solid line) as a function
of the size in pixels of the convolving Gaussian for three different
choices of the original mean number of counts per pixel ($\mu$). On
the left, $\mu = 0.01$ counts pixel$^{-1}$. In the centre, $\mu = 0.1$
counts pixel$^{-1}$. On the right, $\mu = 1.0$ counts
pixel$^{-1}$. Plotted for comparison (dashed line) is the level that
would be correct if the statistics were Gaussian (mean plus 3 times
the standard deviation). It will be seen that the two are increasingly
similar for large convolving Gaussian and large $\mu$, as
expected. Dotted lines in the left-hand figure show the effects of
assuming that convolving with a Gaussian is equivalent to binning with
a bin size equal to the effective area of the Gaussian; they represent
the upper and lower bounds from Poisson statistics on the number of
counts corresponding to the 99.865 per cent confidence level. As
discussed in the text, this approach systematically overestimates the
errors.}
\label{gauss}
\end{figure*}
\clearpage

\begin{thebibliography}{}
\bibitem[]{12}Andernach H., Feretti L., Giovannini G., Klein U., Rosetti E., Schnaubelt J., 1992, A\&AS, 93, 331

\bibitem[]{27}Balcells M., Morganti R., Oosterloo T., P\'erez-Fournon I., Gonz\'alez-Serrano J.I., 1995, A\&A, 302, 665

\bibitem[]{55}Bicknell G.V., 1994, ApJ, 422, 542

\bibitem[]{67}Birkinshaw M., Laing R.A., Peacock J.A., 1981, MNRAS, 197, 253

\bibitem[]{68}Birkinshaw M., Worrall D.M., 1993, ApJ, 412, 568

\bibitem[]{80}B\"ohringer H., Voges W., Fabian A.C., Edge A.C., Neumann D.M., 1993, MNRAS, 264, L25

\bibitem[]{125}Carilli C.L., Perley R.A., Harris D.E., 1994, MNRAS, 270, 173

\bibitem[]{131}Cioffi D.F., Jones T.W., 1980, AJ, 85, 368

\bibitem[]{148}Cornwell T., Perley R., 1984, in Bridle A.H., Eilek J.A., eds, Physics of Energy Transport in Radio Galaxies, NRAO Workshop no.~9, NRAO, Green Bank, West Virginia, p.~39

\bibitem[]{191}Fanaroff B.L., Riley J.M., 1974, MNRAS, 167, 31P

\bibitem[]{198}Feigelson E.D., Laurent-Muehleisen S.A., Kollgaard R.I., Fomalont E., 1995, ApJ, 449, L149

\bibitem[]{203}Feretti L., Perola G.C., Fanti R., 1992, A\&A, 265, 9

\bibitem[]{243}Hardcastle M.J., Alexander P., Pooley G.G., Riley J.M., 1996, MNRAS, 278, 273

\bibitem[]{245}Hardcastle M.J., Alexander P., Pooley G.G., Riley J.M., 1997, MNRAS, 288, L1

\bibitem[]{246}Hardcastle M.J., Birkinshaw M., Worrall D.M., 1998, to
appear in MNRAS (astro-ph/9709228)

\bibitem[]{247}Hardee P.E., Cooper M.A., Clarke D.A., 1994, ApJ, 424, 126

\bibitem[]{282}J\"agers W.J., 1987, A\&AS, 71, 75

\bibitem[]{307}Killeen N.E.B., Bicknell G.V., Ekers R.D., 1988, ApJ, 325, 180

\bibitem[]{330}Laing R.A., 1984, in Bridle A.H., Eilek J.A., eds, Physics of Energy Transport in Radio Galaxies, NRAO Workshop no.~9, NRAO, Green Bank, West Virginia, p.~90

\bibitem[]{334}Laing R.A., 1996, in Hardee P.E., Bridle A.H., Zensus J.A., eds, Energy Transport in Radio Galaxies and Quasars, ASP Conference Series vol.~100, San Francisco, p.~241

\bibitem[]{351}Leahy J.P., Bridle A.H., Strom R.G., 1997, Internet WWW page, at URL: $<$http://www.jb.man.ac.uk/atlas/$>$

\bibitem[]{383}Lupton R.H., Gott J.R., 1982, ApJ, 255, 408

\bibitem[]{406}Miley G.K., Norman C., Silk J., Fabbiano G., 1983, A\&A, 122, 330

\bibitem[]{414}Morganti R., Fanti R., Gioia I.M., Harris D.E., Parma P., de~Ruiter H., 1988, A\&A, 189, 11

\bibitem[]{472}Perley R.A., Willis A.G., Scott J.S., 1979, Nat, 281, 437

\bibitem[]{524}Sarazin C.L., 1986, Rev.\ Mod.\ Phys.\ , 58, 1

\bibitem[]{555}Snowden S., 1997, Internet WWW page, at URL $<$ftp://legacy.gsfc.nasa.gov/rosat/software/fortran/sxrb/README$>$

\bibitem[]{565}Stark A.A., Gammie C.F., Wilson R.W., Bally J., Linke R.A., Heiles C., Hurwitz M., 1992, ApJS, 79, 77

\bibitem[]{626}Worrall D.M., Birkinshaw M., 1994, ApJ, 427, 134

\bibitem[]{627}Worrall D.M., Birkinshaw M., Cameron R.A., 1995, ApJ, 449, 93

\bibitem[]{631}Wyndham J.D., 1966, ApJ, 144, 459

\end{thebibliography}
\end{document}